\documentclass[%
reprint,
  amsmath,amssymb,
  aps,
]{revtex4-1}

\usepackage{graphicx}
\usepackage{dcolumn}
\usepackage{bm}

\usepackage{hyperref}

\begin{document}


\title{$\alpha$-decay properties of even-even superheavy nuclei}

\author{O.N.Ghodsi}

\author{M.Hassanzad}

\affiliation{Department of Physics, Faculty of Basic Sciences,\\
University of Mazandaran, P.O.Box 47416-416, Babolsar, Iran\\}

\begin{abstract}

In this paper, we have calculated the $\alpha$-decay half-lives of superheavy nuclei with $106 \leq Z \leq 126$ and a neutron number of $150 \leq N \leq 200$ within proximity potentials and deformed-spherical Coulomb potentials by using $WS4$ $\alpha$-decay energy and the semi-classical Wentzel-Kramers-Brillouin approximation for penetration probability. Besides, we have included the preformation factor within the cluster-formation model. We have investigated magic numbers and submagic numbers in the mentioned region; with high probability 162, 178, and 184 are predicted as neutron magic numbers. We also have confirmed that there is good agreement between our predicted half-lives and the ones obtained from semiempirical relationships such as Royer, VSS, UDL, and SemFIS2.


\end{abstract}

\maketitle



\section{\label{sec:level1}INTRODUCTION}

In recent years, the subject of superheavy nuclei (SHN) has attracted scrutiny in the field of nuclear physics from both the theoreticians and the experimentalists. Theoreticians \cite{0000,0010,0020} have been attempting to predict the decay properties. On the other hand, Several experiments have been performed \cite{00301,0030,0040,0050,0060} with the purpose of synthesizing SHN.

Because the stability of nuclei is due primarily to the shell effects, the predictions for the shell structures are significant \cite{0070,0080,0090,0091}. The most dominant decay in the superheavy region is $\alpha$-decay. Many studies have been done in this region within different models that use $\alpha$-decay half-lives \cite{0100,0110,0120,0130,0140,0150,0160,0170,0180,0190,0200} to achieve information related to nuclear shell structure. Also, researchers have investigated the island of stability through empirical formulas \cite{0210,0220,0221}.

A study in the superheavy region with Z=110–118 has indicated that the most stable configurations are the deformed ones at neutron number N=162 \cite{0230}. Also in another study \cite{0240}, a doubly magic nucleus was predicted with Z=108 and N=162. A prior paper \cite{0250} was predicted the doubly magic nucleus with Z=114 and N=184, and other investigations \cite{0260,0270,0280,0290} have confirmed these results. More recent studies like Yukawa plus the exponential model with Woods–Saxon single-particle potentials \cite{0300,0310} have also predicted $^{298}$114 as being the next spherical doubly magic nucleus. Non-relativistic microscopic models such as the Skyrme–Hartree–Fock–Bogoliubov method have predicted Z=120 to be as probable as Z=114 to be magic \cite{0320}. Furthermore, Stoyer \cite{0330} has predicted that magic islands exist around Z=120, 124, or 126 and N=184. One of the significant outcomes of these measurements is the increased stability of SHN when approaching N = 184. The heaviest neutron-rich nuclei in the vicinity of the closed spherical shells N=184 and Z=114 (or possibly 120, 122, or 126) were expected to mark a considerable increase in nuclear stability, similar to the effect of closed shells on the stability of the doubly magic $^{208}$Pb nucleus.

The problem of the quantum many-body system is complex; hence, the $\alpha$-preformation factors are obtained from the ratios of calculated to experimental $\alpha$-decay half-lives \cite{0340,0350}. The recently proposed cluster formation model \cite{0360,0370,0380,0390,0400} suggests that the $\alpha$-preformation factor can be extracted in terms of the $\alpha$-cluster formation energy based on the binding energy differences of the participating nuclide. Meanwhile, the behavior of $Q_{\alpha}$ and $P_{\alpha}$ values of 118$\leq$Z$\leq$128 isotopes with increasing neutron number N has been systematically studied \cite{0170}; In consequence, it is suggested that N=178 may be the neutron magic number.

We have already carried out a comprehensive investigation of the SHN \cite{0410}; As a result, we have demonstrated the applicability of nuclear potentials along with a spherical-deformed Coulomb potential for predicting $\alpha$-decay half-lives of SHN, and the present study could be considered as an extension of our earlier works in this region. In this paper, we have performed a considerable study of the $\alpha$-decay of nuclei with 106$ \leq $Z $\leq $126, and $150 \leq N \leq 200$ to find out which possible magic numbers could be placed in which neutron numbers. This paper is organized as follow. The theoretical framework is introduced in Sec.~\ref{sec:level2}. Results and corresponding discussions are given in Sec.~\ref{sec:level3}. The conclusion of the entire work is given in Sec.~\ref{sec:level4}.


\section{\label{sec:level2}THEORETICAL FRAMEWORK}

 \subsection{\label{sub:sub00}Half-life formalism}
 We have calculated the $\alpha$-decay half-life through the formalism
\begin{equation}
{T_{\frac{1}{2}}}={\frac{\ln2}{\nu_{0} P}}.
\end{equation}
Here, $\nu_{0}$ is the assault frequency that is related to the oscillation frequency $\omega$ as
\begin{equation}
{\nu_{0}}={\frac{\omega}{2\pi}}={\frac{(2n_{r}+l+\frac{3}{2})\hbar}{(2\pi\mu R_{n}^2)}}={\frac{(G+\frac{3}{2})}{(1.2\pi\mu R_{0}^2 )}},
\end{equation}
where $R_{n}^2=\frac{3}{5} R_{0}^2$ \cite{0430} and $G=2n_{r}+l$ is the global quantum number \cite{0411}
\begin{equation}
G = 2n_{r} + 1 = \left\{ \begin{array}{lll}
22 & \mbox{for} & N >126            \\
20 & \mbox{for} & 82 < N \leq 126 . \\
18 & \mbox{for} & N\leq82
\end{array}\right.
\end{equation}
The $\alpha$-decay penetration probability $P_{\alpha}$ using the semi classical Wentzel-Kramers-Brillouin (WKB) approximation is defined as
\begin{equation}
P=exp \left( {{-\frac{2}{\hbar}} \int^{r_{b}}_{r_{a}} \sqrt{2\mu(V_{T}(r)-Q_{\alpha})}}dr \right),
\end{equation}
where ${\mu}= m{\frac{A_{\alpha}+A_{d}}{A_{\alpha} A_{d}}}$ is the reduced mass ($A_{\alpha}=4$ and $A_{d}$ is daughter nucleus). Also, $r_{a}$ and $r_{b}$ are the turning points, which are obtained from $V_{T}(r_{a} )=Q_{\alpha}=V_{T} (r_{b})$.

\subsection{\label{sub:sub01}Nuclear potential}
The total interaction potential V$_{T}$(r) between the $\alpha$-particle and the daughter nucleus is taken as follows
\begin{equation}
{V_{T}(r)}={V_{N}(r)+V_{C}(r)+V_{l}(r)}
\end{equation}
where, V$_N$(r), V$_C$(r) and V$_l$(r) are the nuclear potential, the Coulomb potential, and the centrifugal potential, respectively. Because the spin and parity of SHN in the region under study are not known yet, to obtain a precise prediction we neglect the centrifugal potential V$_{l}$(r)  contribution in the total interaction potential. For the calculation of the nuclear potential V$_N$(r), the proximity potential is applied:
\begin{equation}
{V_{N}(r)}={4\pi\gamma b \overline{R}\Phi(\xi)},
\end{equation}
where $\gamma$ is the nuclear surface tension. The details of this formalism are described in Refs.\cite{0420,0430,0440,0450}. In the following, with respect to our previous work, we denote $\gamma$-MS 1967 as prox. 66 and  $\gamma$-PD-LDM 2003 as prox. 03 I.

Using the energy density formalism and Fermi distributions for the nuclear densities Ng$\hat{o}$ 80 and collaborators parameterized the nucleus-nucleus interaction potential in the spirit of proximity concept. The interaction potential can be divided into the geometrical factor and a universal function. The nuclear part of the parameterized potential is defined as \cite{0460}
\begin{equation}
{V_{N}^{Ng\hat{o} 80}(r)}={\bar{R} \phi(r-C_{1}-C_{2})},
\end{equation}
where the nuclear radii $R_{i}$ reads as
\begin{equation}
{R_{i}}={\frac{NR_{ni}+ZR_{pi}}{A_{i}}},    (i=1,2)
\end{equation}
and the equivalent sharp radius for protons and neutrons are given by
\begin{equation}
{R_{(p,n)i}}={r_{{(p,n)i}} A_{i}^{1/3}}.
\end{equation}
Here $r_{pi}$=1.128 fm and $r_{ni}$=$1.1375+1.875\times10^{-4}A_{i}$ fm. The universal function $\phi(r-C_{1}-C_{2})$ is written as
\begin{equation}
\Phi(\xi)= \left\{ \begin{array}{lll}
-33+5.4(s-s_{0})^2                                  & \mbox{for} & s <    s_{0}, \\
-33 exp(\frac{-1}{5}(s-s_{0})^2)                    & \mbox{for} & s \geq s_{0},
\end{array}\right.
\end{equation}
with $ s_{0} =-1.6$ fm. This potential is labeled as Ng$\hat{o}$ 80.

\subsection{\label{sub:sub02}Coulomb potential}
To derive the Coulomb potential for a spherical-deformed nuclear pair, realistic density distributions, and the double-folding model have been used. In this model, the Coulomb interaction potential between spherical-deformed nuclei and deformed-deformed nuclei with separation distance $\vec{R}$ between their centers is given by \cite{0461}
\begin{equation} \label{equ:11}
{V_{C}(\vec{R})}={\int \int d\vec{r_1} d\vec{r_2} \frac{1}{|\vec{s}|} \rho_P (\vec{r}_1) \rho_T (\vec{r}_2)},
\end{equation}
where $\vec{S}=\vec{R}+\vec{r}_1+\vec{r}_2$. $\rho_P$ and $\rho_T$ show the nuclear charge distribution in the projectile and target nuclei which are normalized to the total charge, respectively.
We restrict our derivation to be for the spherical-deformed nuclear pair with the coordinates that are defined as
\begin{equation}
{G(\vec{R},\beta,s)}={\int \rho_T(\vec{R}+\vec{r}) \rho_P(\vec{r}+\vec{s})d\vec{r} },
\end{equation}
where $\beta$ is the orientation angle of the deformed nucleus. After being solving and substituted into Eq. (\ref{equ:11}), $V_{c}(\vec{R},\beta)$ becomes
\begin{widetext}
\begin{equation}
{V_{C}(\vec{R},\beta)}={8 \int_{0}^{\infty} \int_{0}^{\infty} s ds j_0 (ks) k^{2} dk \int d\vec{r} \rho_T(\vec{R}+\vec{r}) j_0 (kr) \int x^{2} dx j_0 (kx)  \rho_P(x) }.
\end{equation}
\end{widetext}
The the charge density distribution of the deformed nucleus is then assumed to be
\begin{equation}
{\rho(r,\theta)}={\frac{\rho_0}{1+e^{\frac{r-R(\theta)}{a}}} },
\end{equation}
where $R(\theta)=r_0 [1+\beta_{2} Y_{20}(\theta,0)+\beta_{4} Y_{40}(\theta,0)]$ is the half density radius of this Fermi distribution. The parameters $\beta_2$ and $\beta_4$ are respectively, the quadrupole and hexadecapole deformation parameters of the residual daughter nucleus and their numerical values are taken from Ref. \cite{0462}.

\subsection{\label{sub:sub03}Cluster-formation model (CFM)}
The cluster–formation model (CFM) is a new quantum mechanical theory was first promoted to calculate the $\alpha$-preformation factors $P_{\alpha}$ of even–even nuclei \cite{0370,0400}. After that, this model was proposed to calculate odd–A and odd–odd nuclei \cite{0360,0380,0381}. The total state $\Psi$ of the parent nucleus is a linear combination of its $n$ possible clusterization states $\Psi_{i}$, which can be defined as
\begin{equation}
{\Psi=\sum_{i=1}^{n}{a_{i}}\Psi_{i}},
\end{equation}
where $a_{i}$ is the superposition coefficient of $\Psi_{i}$. It can be expressed as
\begin{equation}
{a_{i}=\int\Psi_{i}^{*}\Psi d\tau}.
\end{equation}
Orthogonality condition requires that ${\sum_{i=1}^{n}{|a_{i}|^{2}}=1}$. The total wave function is an eigenfunction of the total Hamiltonian $H=\sum_{i=1}^{n}{H_{i}}$, where $H_{i}$ denotes the Hamiltonian for the $i$th clusterization state $\Psi_{i}$. Because all the clusterization states describe the same nucleus, they are assumed to share the same total energy $E$ of the total wave function. So the total energy $E$ can be represented as
\begin{equation}
{E=\sum_{i=1}^{n}{|a_{i}|}^{2}E}={\sum_{i=1}^{n}E_{fi}},
\end{equation}
where $E_{fi}$ is the formation energy of the cluster in the $i$th clusterization state $\Psi_{i}$. Hence, the $\alpha$-preformation factor can be obtained by
\begin{equation}
{P_{\alpha}=|a_{\alpha}|^{2}=\frac{E_{f\alpha}}{E}},
\end{equation}
where $a_{\alpha}$ represents the superposition coefficient of the $\alpha$-clusterization state $\Psi_{\alpha}$, $E_{f \alpha}$ represents the formation energy of the $\alpha$ cluster, and $E$ represents energy actually composed of the formation energy of the $\alpha$ cluster and the interaction energy between the $\alpha$ cluster and the daughter nuclei. Within the CFM, for even–even nuclei, the $\alpha$-cluster–formation energy $E_{f\alpha}$ and the total energy $E$ of a considered system can be expressed as\cite{0391}
\begin{widetext}
\begin{equation}
{E_{f\alpha}=3B(A,Z)+B(A-4,Z-2)-2B(A-1,Z-1)-2B(A-1,Z)},
\end{equation}
\end{widetext}
\begin{equation}
{E=B(A,Z)-B(A-4,Z-2)},
\end{equation}
where $B(A,Z)$ is the binding energy of the nucleus with the mass number A and the proton number Z. In this paper, the data of nuclei binding energies are taken from the latest evaluated atomic mass table or WS4 \cite{0392} for nuclei under calculation .

\subsection{\label{sub:sub04}Semiempirical relationship for $\alpha$ decay}
One of the purposes of this study is to predict the half-life of the for SHN's $\alpha$ decay for which the experimental data of half-life have not been reported yet. Hence, in order to compare our obtained results with other predictions, some semi-experimental relationships used in this work are summarized in the following.

\subsubsection{The Viola-Seaborg-Sobiczewski (VSS) semiempirical relationship}
One of the most famous formulas for calculating alpha decay half-lives is the five-parameter formula offered by Viola and Seaborg\cite{0470}:
\begin{equation}
{\log_{10}(T_{\frac{1}{2}})=(aZ+B) Q^{-\frac{1}{2}} + cZ + D + h_{log}},
\end{equation}
where Z is the atomic number of the parent nucleus and a, b, c and d are $1.66175$, $-8.5166$, $-0.20228$ and $-33.9069$ \cite{0480}, respectively, and
\begin{equation}
h_{log} = \left\{ \begin{array}{llll}
0     & \mbox{for} & Z=even, & N=even, \\
0.772 & \mbox{for} & Z=odd,  & N=even, \\
1.066 & \mbox{for} & Z=even, & N=odd,  \\
1.114 & \mbox{for} & Z=odd,  & N=odd.
\end{array}\right.
\end{equation}

\subsubsection{The analytical formula for $\alpha$-decay half-life (Royer)}
An analytical formula for $\alpha$-decay half-lives has been developed by Royer \cite{0490} and is given by
\begin{equation}
{\log_{10}(T_{\frac{1}{2}})=a+ bA^\frac{1}{6} \sqrt{Z}+\frac{cZ}{\sqrt{Q_{\alpha}}}},
\end{equation}
where A and Z represent the mass and charge number of parent nuclei. The constant a, b, and c are
\begin{widetext}
\begin{equation}
h_{log} = \left\{ \begin{array}{llllll}
a=-25.31 & b=-1.1629 & c=1.5864 & \mbox{for} & Z=even, & N=even, \\
a=-26.65 & b=-1.0859 & c=1.5848 & \mbox{for} & Z=even, & N=odd,  \\
a=-25.68 & b=-1.1423 & c=1.5920 & \mbox{for} & Z=odd,  & N=even, \\
a=-29.48 & b=-1.1130 & c=1.6971 & \mbox{for} & Z=odd,  & N=odd.
\end{array}\right.
\end{equation}
\end{widetext}

\subsubsection{The universal decay law}
A new universal decay law (UDL) for $\alpha$ and cluster decay modes was introduced by Qi \emph{et al.} \cite{0500,0510} as
\begin{equation}
{\log_{10}(T_{\frac{1}{2}})=aZ_{c} Z_{d} \sqrt{\frac{A}{Q_{c}}} +b \sqrt{AZ_{c} Z_{d} (A_{d}^\frac{1}{3} + A_{c}^\frac{1}{3})}+c},
\end{equation}
where $A=\frac{A_{c} A_{d}}{A_{c}+A_{d}}$ and the constant $a=0.4314$, $b=-0.4087$ and $c=-25.7725$ are determined by fitting to experimental data of both $\alpha$ and cluster decays \cite{0500}.

\subsubsection{Semiempirical formula based on fission theory}
Poenaru \emph{et al.} \cite{0511} proposed semiempirical formula for $\alpha$-decay half-lives based on fission theory (SemFIS2), which is expressed as
\begin{equation}
\log_{10}(T_{\frac{1}{2}})=0.43429\chi(x,y)\kappa-20.446+H^{f},
\end{equation}
where
\begin{equation}
{\kappa=2.52956Z_{d}(\frac{A_{d}}{AQ})^{\frac{1}{2}}[arccos\sqrt{r}-\sqrt{r(r-1)}]} ,
\end{equation}
and r=0.423Q(1.5874+$A^{1/3}$)/$Z_{d}$ . The numerical coefficient $\chi$ is a second order polynomial:
\begin{equation}
{\chi(x,y)=B_{1}+x(B_{2}+xB_{4} )+y(B_{3}+yB_{6} )+xyB_{5}}.
\end{equation}
For super heavy fitting, the values are obtained as $B_{1}=0.985415$, $B_{2}=0.102199$, $B_{3}=-0.024863$, $B_{4}=-0.832081$, $B_{5}=1.50572$, and $B_{6}=-0.681221$. The hindrance factor $H^{f}$ is 0.63 for even-odd, 0.51 for odd even, 1.26 for odd-odd and zero for double-even nuclei. The reduced variables x and  y are defined as
\begin{equation}
\left\{ \begin{array}{lll}
x \equiv (N - N_{i})/(N_{i+1} - N_{i})       & \mbox{for} &      N_{i} < N\leq N_{i+1} , \\
y \equiv (Z - Z_{i})/(Z_{i+1} - Z_{i})       & \mbox{for} &      Z_{i} < Z\leq Z_{i+1},
\end{array}\right.
\end{equation}
with $N_{i} = . . . ,51,83,127,185,229, . . . $, $Z_{i} = . . . ,29,51, 83, 127, . . . $; hence for the region of SHN, $x = (N - 127)/(185 - 127)$ and $y = (Z - 83)/(127 - 83)$.

\section{\label{sec:level3}RESULTS AND DISCUSSION}
Since the applicable and exciting subject in nuclear physics is to seek magic numbers and islands of stability in the superheavy region, we have studied the properties of a wide range of nuclei in this area. Their $\alpha$-decay half-lives have been calculated, with regard to the dominant decay mode in this region. For such a purpose, we have applied the formalism introduced in Sec.\ref{sub:sub00}, which has already confirmed a suitable compromise with experimental results \cite{0410}.

To calculate the Coulomb potentials, we have used the formalism of a Coulomb potential for a spherical-deformed nuclear pair. The barrier penetrability of the $\alpha$ particle in a deformed nucleus depends on the orientation of the emitted $\alpha$ particle, with respect to the symmetry axis of the daughter nucleus. The average of penetrability over different directions is done by using the following equation

\begin{equation} \label{equ:31}
{P}={\frac{1}{2} \int_{0}^{\pi} P(Q,\theta) \sin(\theta)d\theta }.
\end{equation}

The total potential including the proximity potential and an exact method to calculate the Coulomb potential between spherical and deformed nuclei in the framework of the double-folding model have been used to determine the penetration probability. It is possible that the Q line does not cross the total potential line in some directions, so the penetration probability and thereby the half-life are not measurable, therefore these directions are neglected in calculation of Eq. (\ref{equ:31}).

In this study, we consider that alpha transitions occur from a parent nuclei in the ground state to a daughter nuclei in the ground state.  Under this condition, all the parent and residual nuclei have zero spin and positive parity. Hence, we are expecting that $\alpha$-decay half-lives of recognizing nuclei are well-suited to exploring shell closure in the superheavy region.

\begin{figure*}
\includegraphics [width=.90\textwidth,origin=c,angle=0] {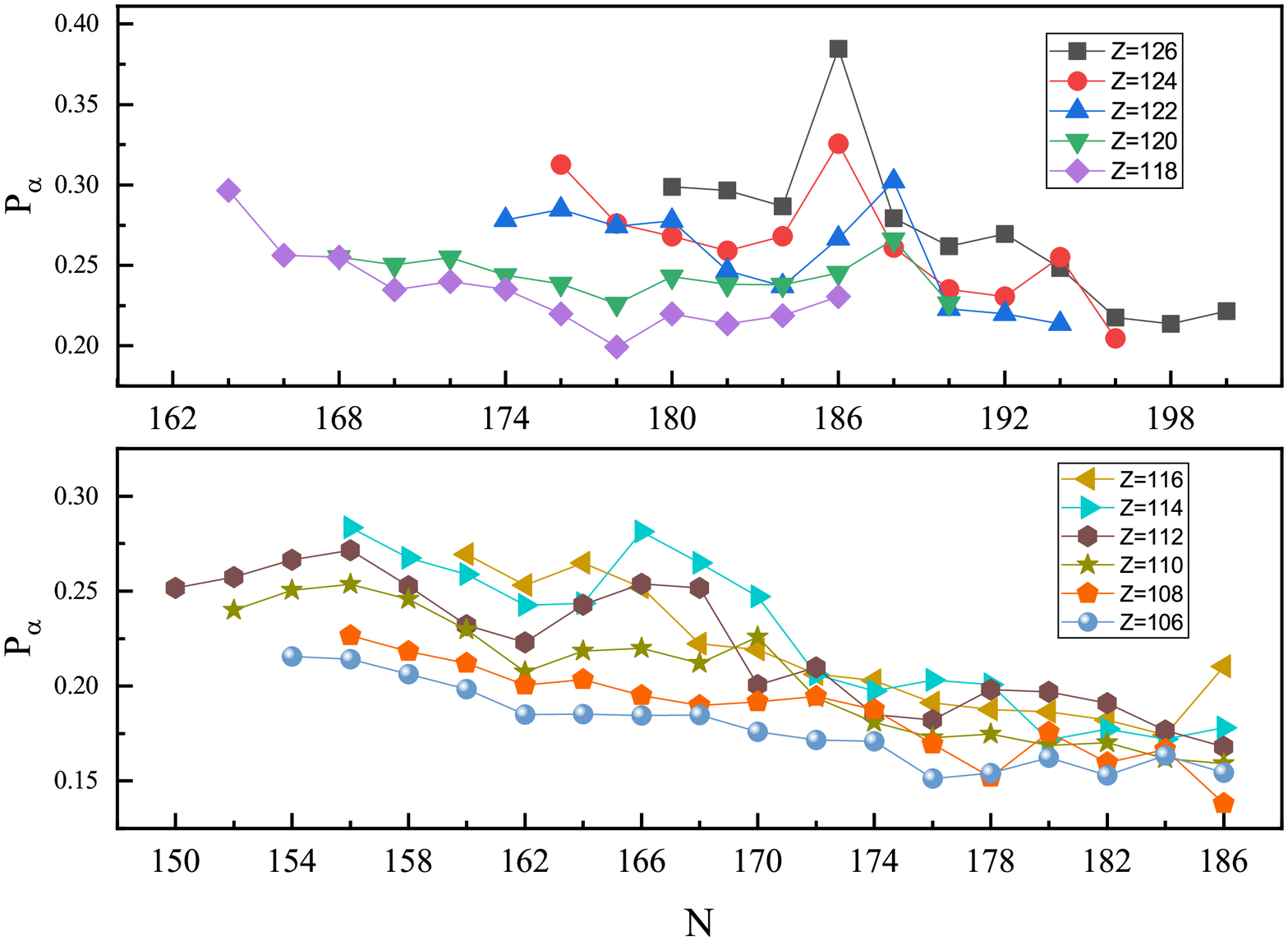} 
\caption{\label{fig:figureP} The $\alpha$-preformation factors ($P_{\alpha}$) with respect to neutron number N for even Z =106 to 126 isotopes that computed within CFM by using $WS4$ mass model.}
\end{figure*}

It is vital to select suitable $\alpha$ decay energies $(Q_{\alpha})$ and preformation probability ($P_{\alpha}$)  values in the calculations because these physical characters significantly affect the accuracy of $\alpha$-decay half-life estimations. Among 20 formalisms, Wang \emph{et al.} \cite{0520}  determined that WS4 \cite{0392} is the formula best suitable to predict the most accurate mass excess and to reproduce the experimental $Q_{\alpha}$ values of the SHN. Moreover, the $\alpha$-particle is assumed to be formed before penetrating the barrier, so we have calculated the  $P_{ \alpha}$ within the CFM by using the WS4 mass model that is explained in Sec.\ref{sub:sub03}. The behavior of $P_{ \alpha}$ is shown in Fig. \ref{fig:figureP}. From this figure, one could find out that the formation of $\alpha$-particles is less probable in some specified neutron numbers. In the following, we have extended our formalism to predict the $\alpha$-decay half-lives of SHN with $106 \leq Z \leq 126$ by using estimated $Q_{\alpha}$ values and also taking preformation factors under consideration to obtain more realistic and reliable predictions in the superheavy region.

To have a quantitative analysis, one can use the root-mean-square deviation (RMSD), which is defined as RMSD=$\sqrt{1/n \sum^{n}_{i=1} [T^{cal}_{i}/T^{Exp.}_{i}]^2}$, in which $n$ is the number of nuclei taken into account and $T^{cal}_{i}$ and $T^{Exp.}_{i}$ are the calculated and experimental $\alpha$-decay half-lives, respectively. For even-even nuclei from Table 3 of Ref. \cite{0410} by considering $P_{\alpha}$ we have obtained RMSDs for prox. 03 I,  prox. 66 and Ngo 80 as 0.5055, 0.5802 and 0.8949, respectively. Furthermore, for the mentioned nuclei with their experimental $Q_{\alpha}$ values, the RMSDs of semiempirical formulas such as SemFIS2, Royer, VSS ,and UDL are 0.52059, 0.4921, 0.4967, and 0.5250, respectively.

\begin{figure*}
\includegraphics [width=1.0\textwidth,origin=c,angle=0] {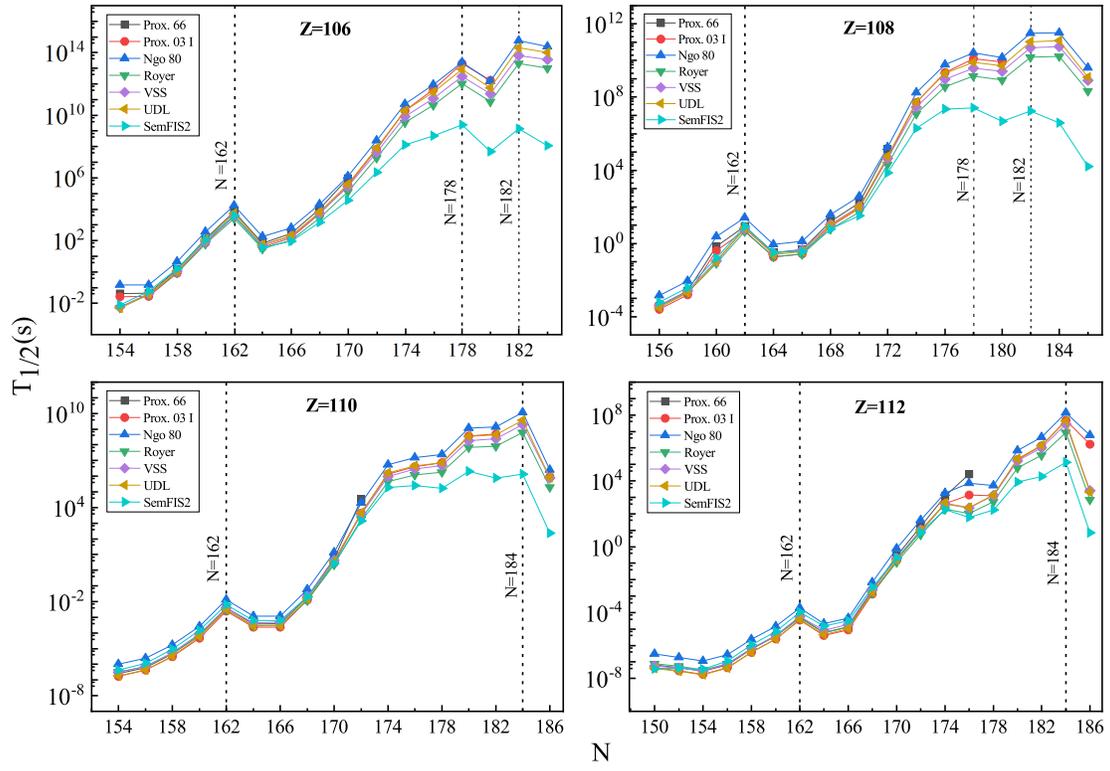} 
\caption{\label{fig:figure1} Logarithmic $\alpha$-decay half-lives with respect to neutron number N, for isotopes of Z$=106, 108, 110, and 112$.}
\end{figure*}

We have estimated the $\alpha$-decay half-lives of SHN and the results are  shown in its logarithmic form in Figs. \ref{fig:figure1} $\_$ \ref{fig:figure3}. From these figures, one can obviously see that half-lives calculated by using our formalisms fundamentally had the same trend. Besides, in the Figs. \ref{fig:figure1} $\_$ \ref{fig:figure3}, there are four additional plotted lines, which refer to the semiempirical relationships for calculating $\alpha$-decay half-lives that were introduced in Sec. \ref{sub:sub04}.

\begin{figure*}
\includegraphics [width=1.0\textwidth,origin=c,angle=0] {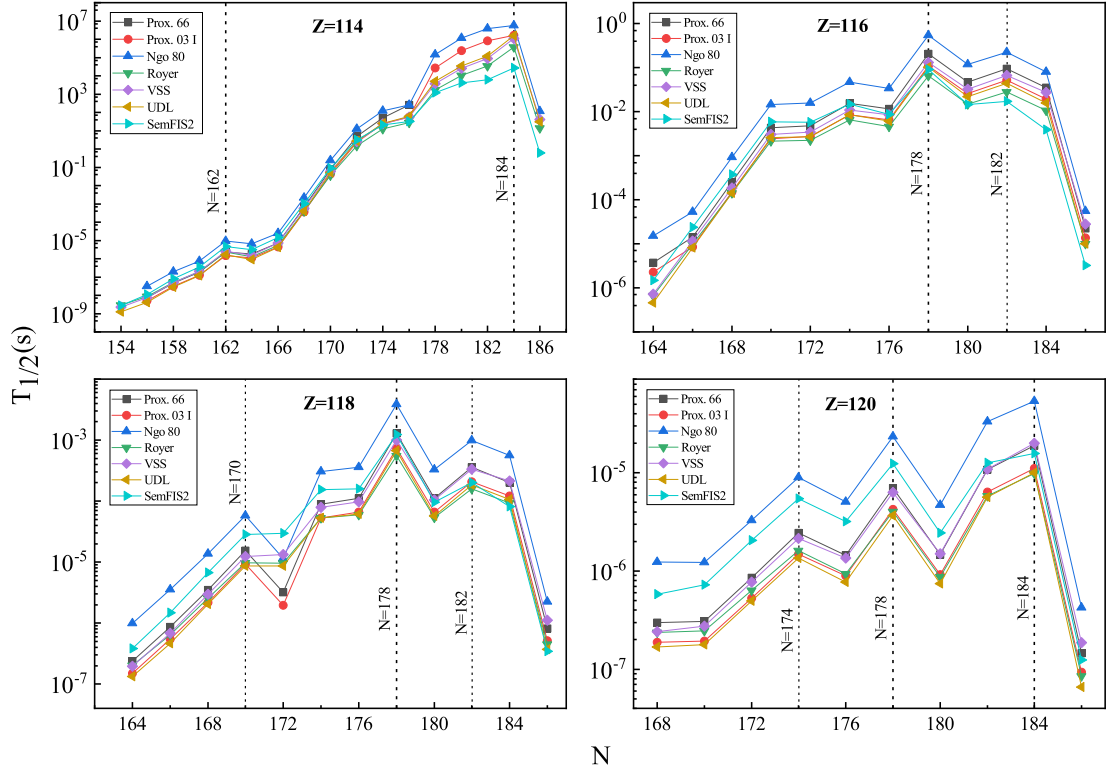} 
\caption{\label{fig:figure2} Logarithmic $\alpha$-decay half-lives with respect to neutron number N, for isotopes of Z$=114, 116, 118, and 120$.}
\end{figure*}

\begin{figure*}
\includegraphics [width=1.0\textwidth,origin=c,angle=0] {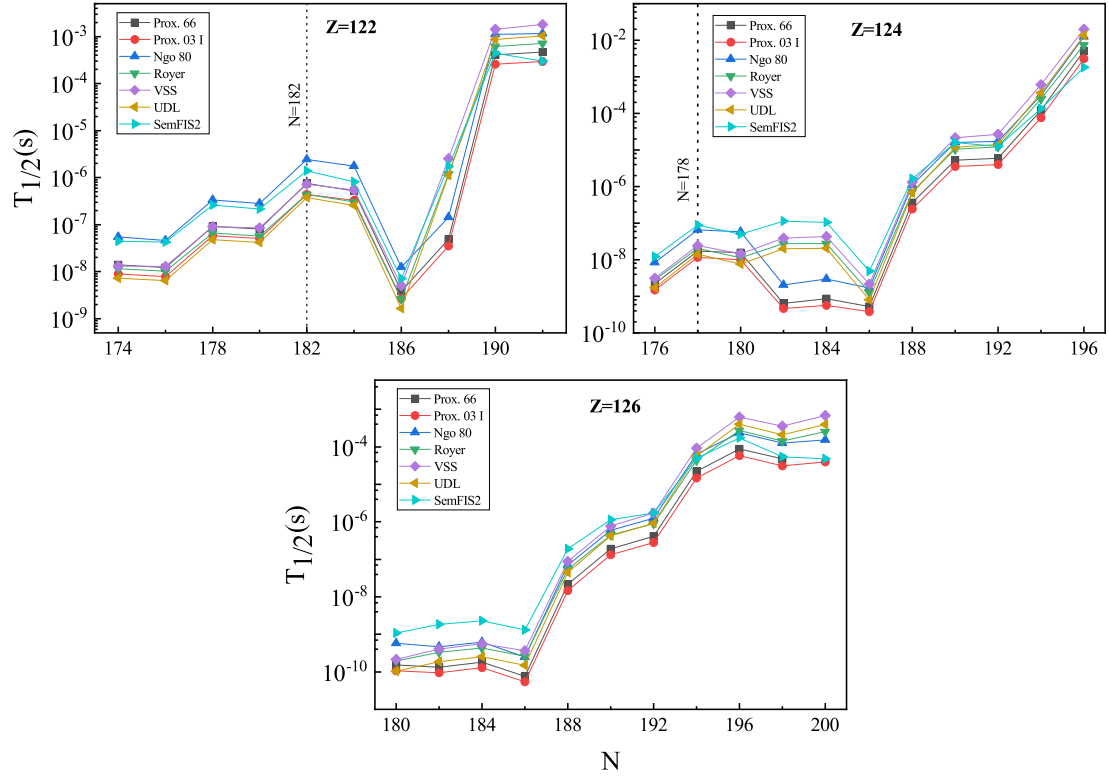} 
\caption{\label{fig:figure3} Logarithmic $\alpha$-decay half-lives with respect to neutron number N, for isotopes of Z$=122, 124, and 126$.}
\end{figure*}

Figure \ref{fig:figure1} determines that for Z=106 and 108 $\alpha$-decay half-lives smoothly increase to N=162 and then dramatically decrease and immediately after that smoothly increase up to N=178. The neutron number N=182, slightly more stable half-lives relative to its very adjacent nuclei. For Z=110 and 112, one is able to notice the same behavior of half-lives up to N=162, after that increase in half-lives continued up to N=184. In Fig \ref{fig:figure2} at Z=114, a likelihood trend near N= 162 is completely recognizable. In consequence, N=162 could be the probable magic number. Further, the second neutron number that shows more stability relative to its neighbor could be N=184.

From Figs.\ref{fig:figure2}, and \ref{fig:figure3} toward  Z=116 to 124, the magic-number-like trend can be observed at N=178, and N=184. According to our preceding claim and the half-lives currently trending around N=182, possibly it is a submagic number. Apparently, one can see that there is a significant maximum in  $P_{\alpha}$ corresponding to the computation of half-lives at N=186; immediately after that, there is a sharp tendency to reach the more stable behavior. This could be an inverse performance concerning the preformation factor, which was explained previously. For Z=126 it seems that we need more proficiency to have a more comprehensive discussion about it.

\section{\label{sec:level4}CONCLUSION}

We have calculated $\alpha$-decay half-lives of a great number of superheavy isotopes with $106 \leq Z \leq 126$ within proximity potentials and Coulomb spherical-deformed potentials. In our formalism, we have used $Q_\alpha$ from the WS4 formalism to compute the penetration probability within the WKB approximation and also we have included the preformation factor within the CFM. Furthermore, we have compared our obtained half-lives with ones computed from semiempirical relationships, such as Royer, VSS, UDL, and SemFIS2 that show a good agreement. Isotopes of different proton numbers generally show the same behavior at N=162, 178, and 184 and in their vicinity, which make them the probable magic numbers in this region. Moreover, at N= 182 and also at 170 and 174, we can expect the submagic numbers. The results we have obtained will prompt inquiries about the nuclear structure and provide information for future experiments.



\newpage

\newpage

\newpage

\newpage

\end{document}